\documentclass[a4paper,12pt]{scrartcl}

\usepackage{amsmath,amssymb,a4,graphicx}
\usepackage{psfrag}
\usepackage{latexsym,amsfonts}

\makeatletter   

\def\section{\@startsection{section}{1}{\z@}{-3.25ex plus -1ex minus
             -.2ex}{1.5ex plus .2ex}{\normalfont\bfseries}}
\def\subsection{\@startsection{subsection}{1}{\z@}{-3.25ex plus -1ex
                minus -.2ex}{1.5ex plus .2ex}{\normalfont\itshape}}

\renewcommand{\title}[1]{\null\vspace{10mm}\noindent
                         {\Large{\bf #1}}\vspace{8mm}}
\newcommand{\authors}[1]{\noindent{\large #1}\vspace{3mm}}
\newcommand{\address}[1]{{\center{\noindent\small\itshape #1\vspace{0mm}}}}

\begin{document}

\begin{titlepage}

\begin{center}

\title{Are there Local Minima in the Magnetic Monopole Potential in Compact QED?}

\authors{H.~Bozkaya$^1$, M.~Faber$^2$, P.~Koppensteiner, M.~Pitschmann$^3$}

\address{$^{1,2,4}$Atominstitut der \"{O}sterreichischen Universit\"{a}ten,
Technische Universit\"at Wien, \\ Wiedner Hauptstra\ss{}e
8--10/142, A-1040 Wien, Austria}
\end{center}
\begin{abstract}
We investigate the influence of the granularity of the lattice on the potential between monopoles.
Using the flux definition of monopoles we introduce their centers of mass and are able to realize continuous 
shifts of the monopole positions. We find periodic deviations from the $1/r$-behavior of the monopole-antimonopole 
potential leading to local extrema. We suppose that these meta-stabilities may influence the order of the phase 
transition in compact QED.
\end{abstract}

\footnotetext[1]{hidir@kph.tuwien.ac.at}

\footnotetext[2]{faber@kph.tuwien.ac.at}

\footnotetext[3]{pitschmann@kph.tuwien.ac.at}

\end{titlepage}

\section{Introduction}

In compact QED (CQED) there are two distinct
phases. The weak coupling regime corresponds to the Coulomb phase
with massless photons and has been investigated in very detail in
the extremely successful program of perturbation theory. However,
there is also a second regime realized in the strong coupling
region, which has become an important research topic since the
1980s. Here, magnetic monopoles cause the creation of electric flux
tubes leading to a linear potential and therefore to confinement.
Although the confinement mechanism in QED seems to be rather different from
the one in QCD, the investigation of the former gives some
inference to the confinement in QCD, which is far more complex to
investigate due to its non-abelian structure. 

In the weak coupling phase the monopoles are supposed to
disappear but they can survive in meta-stable states. There has
been intensive discussion whether the phase transition between
these two phases is of first or second order. For a long time it
seemed that this transition is of first order\cite{JNZ01,B01}. Some later investigations suggested a
second order phase transition\cite{LN01,GNC01,L01,L02}. Recent studies\cite{VD01,ABLS01} 
strongly suggest a first order transition again.

The investigation of the order of the phase transition is of vital importance 
since the continuum limit can only be obtained in the case of a second order transition.

\section{QED on a Space--Time Lattice}

Here we give a short summary of compact QED on the lattice. 
The Wilson action\cite{W01} of QED is given by
\begin{equation}
  S_{W}\, =\, \beta\sum_{x,\mu<\nu}(1 - \cos\theta_{\mu\nu}(x))\,.
\end{equation}
A common choice for the definition of the field-strength is  
\begin{equation}\label{fet}
  ea^{2}F_{\mu\nu}\, =\, \bar{\theta}_{\mu\nu}\,,
\end{equation}
where the physical angle $\bar{\theta}_{\mu\nu}\,\in\,(-\pi,\pi]$ is 
obtained by splitting off $n_{\mu\nu}(x)\,\in\,\{-2,-1,0,1,2\}$ the number of Dirac strings
penetrating a plaquette 
\begin{equation}
  \theta_{\mu\nu}(x)\, =\, \bar{\theta}_{\mu\nu}(x) + 2\pi n_{\mu\nu}(x)\,.
\end{equation}
$\theta_{\mu\nu}(x)$ and $n_{\mu\nu}(x)$ are gauge-dependent contrary to $\bar{\theta}_{\mu\nu}$.
In our simulations we use the definition due to Lang et al.\cite{LR01}
\begin{equation}\label{FT}
  ea^{2}F_{\mu\nu}\, =\, \sin\theta_{\mu\nu}\,,
\end{equation}
where the field strength is a continuous function of $\theta_{\mu\nu}$ and takes into account $2\pi$-periodicity.
This definition is achieved by a variation of the Wilson action\cite{ZFKS01} and is therefore in agreement with 
the Gau{\ss} law on the lattice. As will be described further below, definition (\ref{FT}) allows to investigate 
continuous shifts of monopoles by arbitrary non-integer distances, a property which we need for our investigations.

\section{On Magnetic Charges}

As mentioned in the introduction an interesting feature of $U(1)$
gauge theory is the appearance of magnetic monopoles. So we will shortly recall the fundamental relations
for Dirac monopoles.

Dirac considered a 3-dimensional ball surrounding a magnetic
point charge yielding the following relation 
\begin{equation}\label{Dirac}
  g\, =\, \int_\mathrm{ball}\rho_{m}\,d^{3}\!x\, =\, \int_\mathrm{ball}\vec{\nabla}\vec{B}\,d^{3}\!x\, =\, 
  \oint_\mathrm{sphere}\vec{B}\,d\vec{f}\, =\, \oint_\mathrm{\partial sphere}\vec{A}\,d\vec{s}\,.
\end{equation}
The boundary of a boundary like $\mathrm{\partial sphere}$ is zero, which
implies that the vector potential diverges at a single point on the sphere
for non-vanishing $g$. One is lead to the picture of monopoles
connected by arbitrary lines with a diverging vector potential
along them. These lines are the famous Dirac strings. Their
position is gauge dependent. So, there are
two kinds of divergences. First the "true" physical divergences of
monopoles and the unphysical "gauge" divergences of Dirac strings.

Furthermore, Dirac showed that the electric charge $e$ is
quantized by the existence of a single magnetic charge $g$ since 
the wave function of an electric charge transported along $\mathcal{C}$ is modified by 
\begin{equation}
  \psi \rightarrow \psi\exp\Bigl(-\frac{ie}{\hbar}\int_{\mathcal{C}}\vec{A}\,d\vec{s}\Bigr)
\end{equation}
Transporting $e$ around the boundary $\mathrm{\partial sphere}$ of Eq.~(\ref{Dirac}), the non-ambiguity of $\psi$ 
demands 
\begin{equation}
  \frac{e}{\hbar}\oint_{\partial\mathrm{sphere}}\vec{A}\,d\vec{s}\, =\, \frac{eg}{\hbar}\, =\, 2\pi n\,. 
\end{equation}
Setting $\hbar=1$ yields for the smallest ($n=1$) charge quantum 
\begin{equation}
  e\, =\, \frac{2\pi}{g}\,. 
\end{equation}

\section{Identification of Monopoles}

Dirac monopoles in compact U(1) theory can be identified according to 
DeGrand and Toussaint\cite{DT01} as follows.

The plaquette angle
\begin{equation}
  \theta_{\mu\nu}(x)\, =\, \theta_\mu(x) + \theta_\nu(x+\mu) - \theta_\mu(x+\nu) - \theta_\nu(x) \in (-4\pi,4\pi]
\end{equation}
has no direct physical meaning since the action has a periodicity of $2\pi$ only. 
Furthermore, the sum of six plaquette angles enclosing a cube vanishes
since each link of the cube is counted twice but in opposite 
direction. In the language of forms the cube is the exterior differential 
of the six plaquettes (respectively angles), which are the exterior differentials
of their links. Due to the Bianchi identity ("the boundary of a boundary is zero") the sum of the six plaquette angles of a cube vanishes. 

Dirac strings can be closed or start and end at magnetic monopoles. 
Thus, the magnetic charge in a cube, respectively the number of monopoles in it is given by the sum of in-going minus 
out-going strings. It is an integer multiple of $g = 2\pi/e$. The corresponding
four-current is given by
\begin{equation}\label{DGT}
  j_{m,\mu}^{DGT}(x)\, =\, \frac{2\pi}{e}\,\varepsilon_{\mu\nu\rho\sigma}\Delta_{\nu}n_{\rho\sigma}(x)\,,
\end{equation}
where we used the (forward) lattice derivative $\Delta_{\nu}f(x)\, =\, f(x + \nu) - f(x)$.
Obviously, this current is source-free ($\Delta_{\mu}j_{m,\mu} =
0$). Identifying monopoles this way allows to count the number of 
monopoles in each cube only. For the identification of monopole positions
inside cubes we have to choose another method.   

This second method, which is due to Ref.~\cite{ZFKS01} detects monopoles by measuring 
the magnetic fluxes leaving cubes. The distribution of the flux 
over six plaquettes surrounding a cube reflects the position of 
the monopole inside. Then, the monopole current is given by the dual Maxwell equations, which read in covariant notation
\begin{eqnarray}
  \partial_{\mu}\tilde{F}_{\mu\nu} &=& -j_{m,\nu} \label{ME}\\
  \partial_{\mu}F_{\mu\nu} &=& 0
\end{eqnarray}
with the dual field strength-tensor $\tilde{F}_{\mu\nu} = \frac{1}{2}\,\varepsilon_{\mu\nu\rho\sigma}\,F_{\rho\sigma}$. 
Using the definition of the dual field strength-tensor and Eqs.~(\ref{FT}) and (\ref{ME}) yields for the magnetic 
four-current on the lattice
\begin{equation}\label{Zach}
  j_{m,\mu}(x)\, =\, \frac{1}{2ea^{3}}\,\varepsilon_{\mu\nu\rho\sigma}\Delta_{\nu}\sin\theta_{\rho\sigma}(x)\,.
\end{equation}
Evidently, this current is also conserved ($\Delta_{\mu}j_{m,\mu}=0$). The magnetic charge in 
the cube at position $x$ is given by
\begin{eqnarray}\label{Q}
  Q_{m}(x) &=& a^3j_{m,4}(x)\, =\, \frac{g}{2\pi}\,\sum_{P\in\partial\mathrm{cube}(x)}\sin\theta_{P}\,.
\end{eqnarray}
In this definition the monopole charge in each cube is not an integer multiple of $g$. Only 
for infinitely separated monopoles investigated in the classical limit, the charge inside a large surface 
enclosing a monopole is an integer multiple of $g$ since $\sin\bar{\theta}_{ij}\rightarrow\bar{\theta}_{ij}$. 
This leads to the picture of a single cube yielding the main contribution to the magnetic charge of the monopole, 
while the neighboring cubes contribute with a few percent. Since this charge 
distribution is independent of the lattice constant monopoles are still "point-like". This definition of monopole 
currents was investigated in detail in Ref.~\cite{ZFKS01}. It was shown that the definition (\ref{Q}) is necessary to 
get agreement with the dual superconductor picture of confinement and the dual London equation\cite{SHB01}.   

Using Eq.~(\ref{Q}) we introduce a monopole's center of mass 
\begin{equation}\label{com}
  \vec{R}_{m}^{(\pm)}(t)\, =\, \frac{\sum_{\vec{r}^{(\pm)}}\vec{r}\,Q_m(\vec{r},t)}{\sum_{\vec{r}^{(\pm)}}Q_m
  (\vec{r},t)}\,.
\end{equation}
depending on the flux distribution around the monopole where $\vec{r}^{(+)}$ corresponds to space-like cubes with a 
positive contribution to the magnetic charge and correspondingly for $\vec{r}^{(-)}$. With definition (\ref{com}) it is 
possible to 
obtain positions of monopoles which are not restricted to integer values. 

\section{Monopole Generation}

For the realization of a classical monopole-configuration on the lattice we discuss two possibilities: 

\begin{itemize}
\item Since a plaquette is part of the surface of four cubes, a non-vanishing plaquette value corresponds to non-zero
magnetic currents in these cubes. On the dual lattice the corresponding links assemble a plaquette.
Following this line of thought, fixing the value of an $xy$-plaquette leads to a rotating current in the 
$zt$-plane. Further fixing of all $xy$-plaquettes with given coordinates $x,y,z$ and arbitrary $t$ and closing them 
periodically at the boundary of the lattice yields two currents in $t$-direction but with opposite sign. For sufficiently 
large plaquette angles the two currents in $t$-direction represent a static monopole-antimonopole pair. Thus, arbitrary 
monopole configurations can be realized by fixing corresponding plaquettes appropriately. 

After specifying the constraints for a static monopole pair we minimize the action by local cooling. 
In fig.~\ref{JTHE} the sum of the absolute values of the magnetic charges in one time slice
\begin{eqnarray}
  ||Q_{m}||\, =\, \sum_{\vec r}|Q_{m}(\vec{r},t=t_0)| \,.
\end{eqnarray}
in dependence on the angle of the plaquette separating the monopole pair is shown. 

\begin{figure}
  \centering
  \psfrag{Angle}{$\theta_{\Box}$}
  \psfrag{J}{$||Q_{m}||/g$}
  \psfrag{ 0}{$0$}
  \psfrag{ 0.5}{$0.5$}
  \psfrag{ 1}{$1$}
  \psfrag{ 1.5}{$1.5$}
  \psfrag{ 2}{$2$}
  \psfrag{0}{$0$}
  \psfrag{pi}{$\pi$}  
  \psfrag{-pi}{$-\pi$}
  \psfrag{2pi}{$2\pi$}
  \psfrag{-2pi}{$-2\pi$}
  \psfrag{measured data}{measured data}
  \psfrag{DeGrand-Toussaint}{DeGrand-Toussaint}
  \psfrag{analytical approximation}{analytical approximation}
  \includegraphics[width=0.8\linewidth]{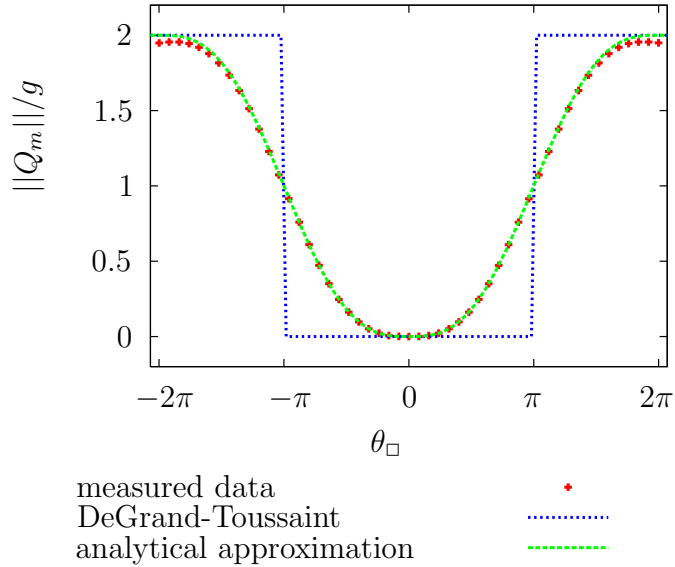}
\caption{Monopole-charge in dependence on the plaquette-angle $\theta_{\Box}$. Besides the
measured data and the result due to the definition of DeGrand-Toussaint an analytical approximation formula is plotted 
which is given by $||Q_{m}||/g=\frac{1}{\pi}|\theta_{\Box} - \sin\theta_{\Box}|$.}
\label{JTHE}  
\end{figure}

Using definition (\ref{DGT}) yields zero charge for plaquette angles in the interval $(-\pi,\pi]$ followed by a 
discrete jump at $\pm\pi$ to $2g$ when a Dirac string penetrates the plaquette. Using definition (\ref{Zach}) this 
jump is spread out (see fig.~\ref{JTHE}). This is caused by the fact that the charge of the monopole in definition 
(\ref{Zach}) is distributed over several cubes, allowing for charge annihilation for a monopole pair 
at close distance. Furthermore, values for the approximation formula 
\begin{eqnarray}
 ||Q_{m}||\,=\,\frac{g}{\pi}\,|\theta_{\Box} - \sin\theta_{\Box}|    
\end{eqnarray}
are plotted in fig.~\ref{JTHE}. This expression is obtained by considering two large closed surfaces each surrounding 
one monopole and assuming that all plaquettes $P$, except the tuned one $\Box$, have small values $\theta_{P}$.
\\
\item Constraints with local chemical potentials $\mu$ introduced at single cubes can be used to create monopoles at 
given positions. The potentials are introduced in the action as follows
\begin{equation}\label{SGMU}
  S_{\mu}\, =\, \sum_{P}(1 - \cos\theta_{P}) + \sum_{C}\mu(C)J(C)\,,
\end{equation}
with the plaquette angle $\theta_{P}$, the chemical potential $\mu(C)$, the magnetic current $J(C)$
and the sum $C$ extending over all cubes. Fixing a chemical potential $\pm|\mu|$ for a static monopole pair 
at a distance of one lattice spacing followed by local cooling preserves such a monopole-antimonopole pair 
for values $|\mu|>2.3$ only, see 
fig.~\ref{JMU}.

\begin{figure}
  \centering
  \psfrag{chp}{$|\mu|$}
  \psfrag{Q}{$||Q_m||/g$}
  \psfrag{ 0}{$0$}
  \psfrag{ 0.5}{$0.5$}
  \psfrag{ 1}{$1$}
  \psfrag{ 1.5}{$1.5$}
  \psfrag{ 2}{$2$}
  \psfrag{ 3}{$3$}
  \psfrag{ 4}{$4$}
  \psfrag{ 5}{$5$}
  \psfrag{ 6}{$6$}
  \includegraphics[width=0.8\linewidth]{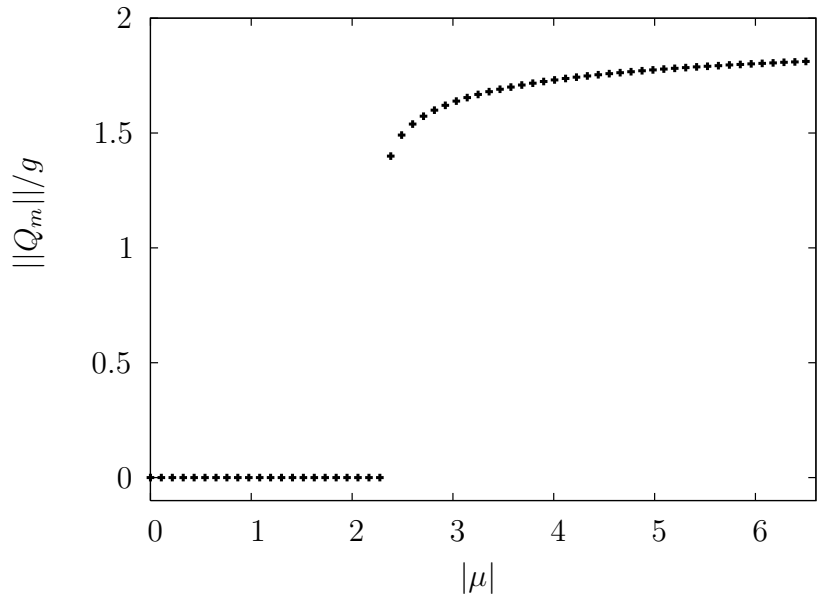}
\caption{Absolute value of magnetic charge in dependence on the absolute value of the 
chemical potential $|\mu|$ introduced at neighboring cubes.
}\label{JMU} 
\end{figure}

One can observe that the charge for the monopole pair created at $|\mu|\approx2.3$ is approximately $1.5g$ and increases
continuously to a value near $2g$. For smaller values of $|\mu|$ the chemical potential is not strong enough to keep apart a 
monopole pair at a distance of one against their magnetic attraction. The monopoles approach each other 
with decreasing $|\mu|$, therefore the magnetic charge densities of monopoles and antimonopoles overlap leading to 
gradual annihilation. For increasing $|\mu|$ the monopoles are
closer at the centers of the cubes and their overlap decreases. Since the charge of the monopole is distributed over more 
than a single cube a value of $2g$ cannot be attained for a monopole pair at distance one. In a strict sense, only 
infinitely separated monopoles obtain the value $2g$. 
Furthermore, as soon as a monopole pair is created one would expect local minima of the action $S_{\mu}$. Indeed 
this is true for $|\mu|>2.3$ as can be seen in fig.~\ref{SETHE}. 

\begin{figure}
  \centering
  \psfrag{theta}{$\theta_{\Box}$}
  \psfrag{S}{$S_\mu$}
  \psfrag{mu=10}{$|\mu|$=1}
  \psfrag{mu=15}{$|\mu|$=1.5}
  \psfrag{mu=20}{$|\mu|$=2}
  \psfrag{mu=25}{$|\mu|$=2.5}
  \psfrag{mu=30}{$|\mu|$=3}
  \psfrag{mu=35}{$|\mu|$=3.5}
  \psfrag{mu=40}{$|\mu|$=4}
  \psfrag{mu=45}{$|\mu|$=4.5}
  \psfrag{mu=50}{$|\mu|$=5}
  \psfrag{0}{$0$}
  \psfrag{pi/2}{$\pi/2$}
  \psfrag{pi}{$\pi$}
  \psfrag{3pi/2}{$3\pi/2$}
  \psfrag{2pi}{$2\pi$}
  \psfrag{-2}[cr][cr]{$-2$}
  \psfrag{ 0}{$0$}
  \psfrag{ 2}{$2$}
  \psfrag{ 4}{$4$}
  \psfrag{ 6}{$6$}
  \psfrag{ 8}{$8$}
  \psfrag{ 10}{$10$}
  \psfrag{ 12}{$12$}
  \includegraphics[width=0.8\linewidth]{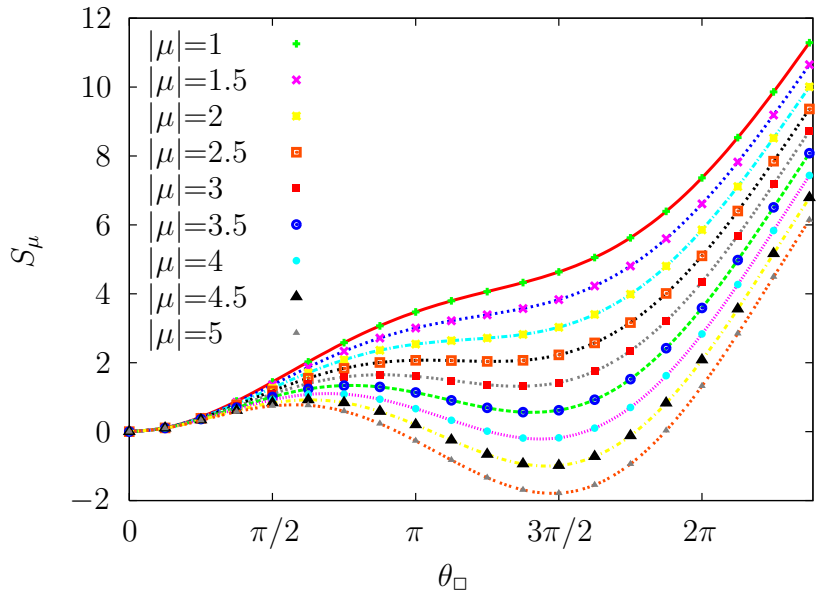}
  \caption{$S_{\mu}$ according to Eq.~(\ref{SGMU}) for a monopole pair
    in neighboring cubes in dependence on the phase $\theta_{\Box}$ of
    the plaquette separating the monopoles for various values of
    $|\mu|$.}
\label{SETHE} 
\end{figure}

\end{itemize}
In fig.~\ref{SETHE} the action $S_{\mu}$ in dependence on the value $\theta_{\Box}$
of the plaquette, which separates the monopole pair is shown for various values of $|\mu|$. One can clearly recognize 
local minima of $S_{\mu}$ for $|\mu|>2.3$.     

\section{The Monopole-Antimonopole Potential in the Classical Limit}

The potential $V$ of a static monopole pair in the classical limit corresponds to the action $S\propto V-T$, since the kinetic 
energy $T$ vanishes. Using the monopole definition (\ref{DGT}) due to DeGrand-Toussaint\cite{DT01} the monopole charge 
is contained entirely in single cubes and the potential can be determined for integer distances only. Due to the
electro-magnetic duality a monopole pair feels a $1/r$-potential at large separations. 

\begin{figure}
  \centering
  \psfrag{monopole position x}{monopole position $x$}
  \psfrag{Angle}{$\theta_{\Box}$}
  \psfrag{antimonopole}{antimonopole}
  \psfrag{monopole}{monopole}
  \psfrag{plaquette}{plaquette}
  \psfrag{pi}{$\pi$}
  \psfrag{pi/2}{$\pi/2$}
  \psfrag{-0.5}{$-0.5$}
  \psfrag{ 0}{$0$}
  \psfrag{ 0.5}{$0.5$}
  \psfrag{ 1}{$1$}
  \psfrag{ 1.5}{$1.5$}
  \psfrag{ 2}{$2$}
  \includegraphics[width=0.8\linewidth]{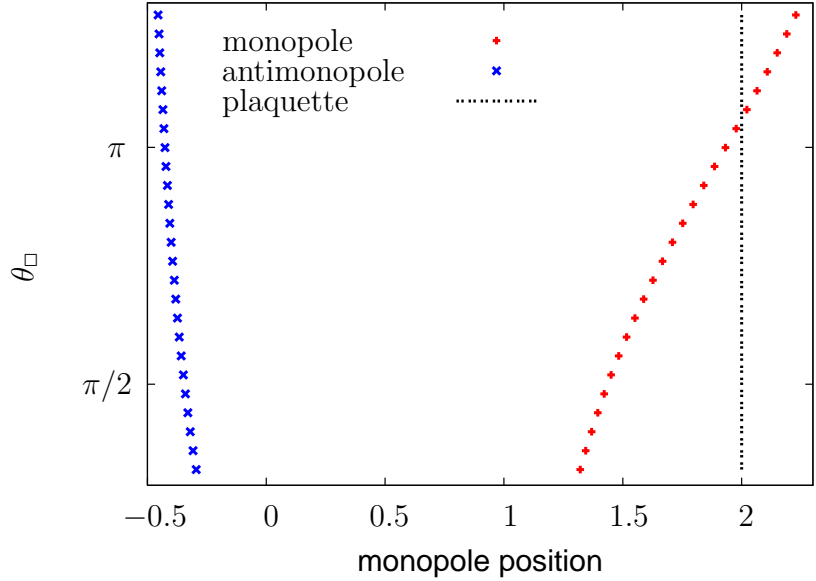}
\caption{Relation between $\theta_{\Box}$ and the positions of monopoles. The vertical 
straight line at $x=2$ indicates the position of the tuned plaquette $\theta_{\Box}$. Introducing a chemical potential 
$\mu=-2.5$ at $x=-0.5$ and tuning of $\theta_{\Box}$ followed by cooling leads to an antimonopole in the cube at $x=-0.5$ 
and a monopole in the vicinity of the tuned plaquette.}\label{moving}
\end{figure}

We investigate the monopole-antimonopole potential not only for integer distances but for arbitrary real 
ones. For this purpose monopoles must be movable inside cubes. This can be achieved by tuning of appropriate 
plaquettes. Fig.~\ref{moving} shows the result of such tuning. 

The chemical potential is set to $\mu=-2.5$ at the cube centered at $(-0.5,0.5,0.5)$, while the angle $\theta_{\Box}$ 
of the plaquette with its center at $(2,0.5,0.5)$ is tuned. After choosing appropriate values of $\theta_{\Box}$ cooling 
generates an antimonopole near $(-0.5,0.5,0.5)$ the center of the cube 
with non-vanishing $\mu$. The position of the monopole depends strongly on the value of $\theta_{\Box}$. For 
$\theta_{\Box}>\pi$ the plaquette is pierced by a Dirac-string and the position of the monopole is larger than two. 
With decreasing $\theta_{\Box}$ the monopole is continuously shifted towards the antimonopole. One can observe 
that for $\theta_{\Box}=\pi$ the monopole is shifted slightly to an $x$-value below $2$, which is due to magnetic 
attraction. Calculations for large separations with and without chemical potentials show that the Wilson part of the 
action $S_{\mu}$ does not differ. 
  
\begin{figure}
  \centering
  \psfrag{distance}{distance}
  \psfrag{V}{$V$}
  \psfrag{0}{$0$}
  \psfrag{1}{$1$}
  \psfrag{2}{$2$}
  \psfrag{3}{$3$}
  \psfrag{4}{$4$}
  \psfrag{5}{$5$}
  \psfrag{ 0}{$0$}
  \psfrag{ 1}{$1$}
  \psfrag{ 2}{$2$}
  \psfrag{ 3}{$3$}
  \psfrag{ 4}{$4$}
  \psfrag{ 5}{$5$}
  \psfrag{ 6}{$6$}
  \psfrag{ 8}{$8$}
  \psfrag{ 10}{$10$}
  \hspace{10mm}\includegraphics[width=0.8\linewidth]{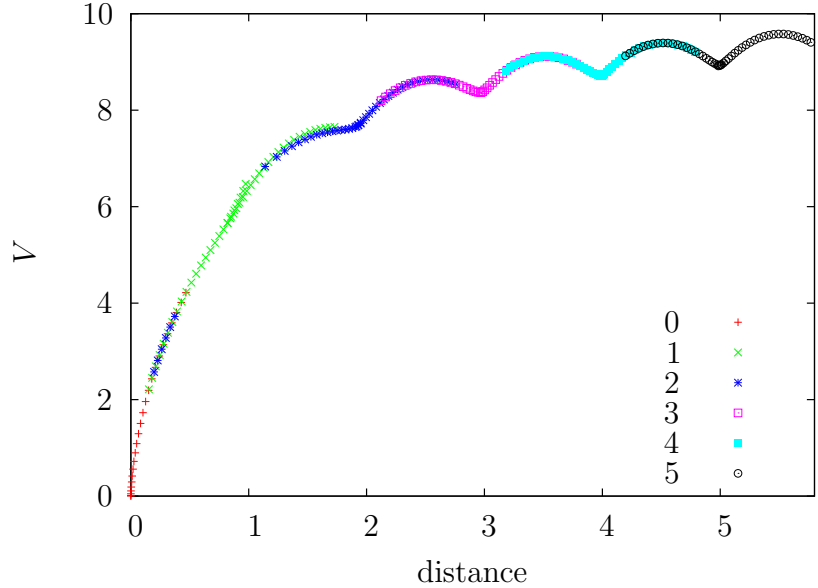}
\caption{The monopole-potential for arbitrary separations is shown. The different 
symbols indicate the position $x_{\Box}$ of the tuned plaquette which allows to vary the monopole position.}\label{action}
\end{figure}

Using the above procedure it is possible to generate monopole pair configurations at arbitrary real distances. 
The potential of a static monopole pair in dependence on their distance yields fig.~\ref{action}.

The coordinate $x_{\Box}$ of the tuned plaquette $\theta_{\Box}$ varies from $1$ to $5$ in integer steps. 
The values of $x_{\Box}$ used 
to determine the potential are indicated in the figure. It is interesting to observe that the well-known $1/r$-potential 
is superimposed by oscillations with a wave length of lattice spacing $a$. These oscillations are obviously due to 
the granularity of the lattice. Local minima are located at integer distances, while the maxima can be found at 
half-integer separations.

This behavior of the potential at non-integer distances may lead to the well-known observation that some monopole loops 
survive the transition from the confined to the Coulomb phase. Refs.~\cite{VD01} and \cite{ABLS01} report strong 
indications that this phase transition is of first order. A vanishing of the two-state signal was detected in 
Refs.~\cite{LN02},\cite{JLN01} on a lattice with a topology homo-topic to an S$^4$-sphere. This was explained by the
fact that for such a topology monopole loops may contract and disappear without restriction. A discretization of an S$^4$ 
where the curvature is distributed almost homogeneously over the lattice favors a contraction of arbitrary monopole loops
and decreases the influence of meta-stabilities like those appearing in Fig.~\ref{action}.

\section{Conclusion and Outlook}

Magnetic monopoles defined by their charge distributions can take arbitrary positions on a discrete space-time lattice. 
With this definition the potential can be calculated for arbitrary non-integer distances 
between monopoles. Local chemical potentials provide an efficient tool to create monopoles on the lattice 
and to realize arbitrary monopole configurations. Magnetic monopoles feel a Coulomb potential in analogy to electric 
point-like charges, if they are located in the centers of the lattice cubes. The above results demonstrate deviations 
from the $1/r$-behavior when the monopoles are shifted away from the centers. We find it interesting to investigate 
whether the meta-stabilities appearing in the potential due to the resulting hysteresis are strong enough to influence 
the order of the phase transition. Methods to remove the influence of the granularity of the lattice will be a subject of 
our further investigations.

\section*{Acknowledgments}

This work was supported in part by ``Fonds zur F\"orderung der Wissenschaften'' (FWF) under contract 
P16910-N12 (P.K., M.P.) and by ``Fonds zur F\"orderung der Wissenschaften'' (FWF) under contract 
P15015-N08 (H.B., M.P.).


\begin{thebibliography}{99}
\bibitem{JNZ01}J.~Jers\'ak, T.~Neuhaus and P.~M.~Zerwas,
        Phys. Lett. B {\bf 133}, 103 (1983).
\bibitem{B01}G.~Bhanot,
        Nucl. Phys. B {\bf 205}, 168 (1982).
\bibitem{LN01}B.~Lautrup and M.~Nauenberg,
        Phys. Lett. B {\bf 95}, 63 (1980).
\bibitem{GNC01}R.~Gupta, M.~A.~Novotny, and R.~Cordery,
        Phys. Lett. B {\bf 172}, 86 (1986).
\bibitem{L01}C.~B.~Lang,
        Phys. Rev. Lett. {\bf 57}, 1828 (1986).
\bibitem{L02}C.~B.~Lang,
        Nucl. Phys. B {\bf 280}, 255 (1987).
\bibitem{VD01}M.~Vettorazzo and P.~de Forcrand,
        Nucl. Phys. B {\bf 686}, 85 (2004),
        [arXiv:hep-lat/0311006].
\bibitem{ABLS01}G.~Arnold, B.~Bunk, T.~Lippert and K.~Schilling,
        Nucl. Phys. Proc. Suppl. {\bf 119}, 864 (2003),
        [arXiv:hep-lat/0210010].
\bibitem{W01}K.~G.~Wilson,
        Phys. Rev. D {\bf 10}, 2445 (1974).
\bibitem{LR01}C.~B.~Lang and C.~Rebbi,
        Phys. Rev. D {\bf 35}, 2510 (1987).
\bibitem{ZFKS01}M.~Zach, M.~Faber, W.~Kainz, P.~Skala,
        Phys. Lett. B {\bf 358}, 325 (1995),
        [arXiv:hep-lat/9508017].
\bibitem{SHB01}V.~Singh, R.~H.~Haymaker, D.~A.~Browne,
        Phys. Rev. D {\bf 47}, 1715 (1993). 
        [arXiv:hep-lat/9206019]
\bibitem{DT01}T.~DeGrand and D.~Toussaint,
        Phys. Rev. D {\bf 22}, 2478 (1980).
\bibitem{LN02}C.~B.~Lang, T.~Neuhaus,
        Nucl. Phys. Proc. Suppl. {\bf 34}, 543 (1994),
        [arXiv:hep-lat/9311030].
\bibitem{JLN01}J.~Jersak, C.~B.~Lang, T.~Neuhaus,
        Phys. Rev. D {\bf 54}, 6909 (1996),
        [arXiv:hep-lat/9606013].
\end{thebibliography}
\end{document}